\documentclass[a4paper]{iopart}

\usepackage{enumitem}
\usepackage{graphicx}
\usepackage{color}

\usepackage[pdftex,colorlinks=true,bookmarks=false,citecolor=blue,urlcolor=blue]{hyperref} 
\usepackage[T1]{fontenc} 

\begin{document}

\title{GEO\,600 and the GEO-HF upgrade program: successes and challenges}

\author{K~L~Dooley$^{1, 10}$, 
  J~R~Leong$^1$, T~Adams$^{3,4}$, C~Affeldt$^1$, A~Bisht$^1$,
  C~Bogan$^1$, J~Degallaix$^5$, C~Gr\"{a}f$^{1,2}$, S~Hild$^2$, J~Hough$^2$,
  A~Khalaidovski$^1$, N~Lastzka$^1$,
  J~Lough$^1$, H~L\"uck$^1$, D~Macleod$^{3,6}$, L~Nuttall$^{3,7}$, M~Prijatelj$^{1,8}$,
  R~Schnabel$^1$, E~Schreiber$^1$, J~Slutsky$^{1,9}$, B~Sorazu$^2$,
  K~A~Strain$^2$, H~Vahlbruch$^1$, M~W\k{a}s$^{1,4}$, B~Willke$^1$, H~Wittel$^1$, 
  K~Danzmann$^1$, and H~Grote$^1$}
\address{$^1$ Max-Planck-Institut f\"ur Gravitationsphysik
(Albert-Einstein-Institut) und Leibniz Universit\"at Hannover, Callinstr.\ 38,
D-30167 Hannover, Germany}
\address{$^2$ SUPA, School of Physics and Astronomy, The University of Glasgow,
  Glasgow, G12 8QQ, UK}
\address{$^3$ Cardiff University, Cardiff CF24 3AA, United Kingdom}
\address{$^4$ Laboratoire d'Annecy-le-Vieux de Physique des Particules (LAPP),
Universit\'{e} de Savoie, CNRS/IN2P3, F-74941 Annecy-le-Vieux, France}
\address{$^5$ Laboratoire des Mat\'{e}riaux Avanc\'{e}s (LMA), IN2P3/CNRS,
Universit\'{e} de Lyon, F-69622 Villeurbanne, France}
\address {$^6$ Louisiana State University, Baton Rouge, LA 70803, USA }
\address {$^7$ Syracuse University, Syracuse, NY 13244, USA }
\address{$^8$ European Gravitational Observatory (EGO), I-56021 Cascina (Pi), Italy}
\address{$^9$ CRESST and Gravitational Astrophysics Laboratory NASA/GSFC,
  Greenbelt, MD 20771, USA}
\address{$^{10}$ The University of Mississippi, Oxford, MS 38677, USA}

\ead{kldooley@olemiss.edu}

\date{\today}

\begin{abstract}
The German-British laser-interferometric gravitational wave detector
GEO\,600 is in its 14th year of operation since its first lock in
2001. After GEO\,600 participated in science runs with other first-generation
detectors, a program known as GEO-HF began in 2009. The goal
was to improve the detector sensitivity at high frequencies,
around 1\,kHz and above,
with technologically advanced yet minimally invasive
upgrades. Simultaneously, the detector would record science quality data in between commissioning
activities. As of early 2014, all of the planned upgrades have been
carried out and sensitivity improvements of up to a factor of four
at the high-frequency end of the observation band have been achieved. Besides science data collection,
an experimental program is ongoing with the goal to further improve the
sensitivity and evaluate future detector technologies.  
We summarize the results of the
GEO-HF program to date and discuss its successes and challenges.
\end{abstract}


\section{Introduction}
\begin{figure}[tb]
\centering
\includegraphics[width=0.9\columnwidth]{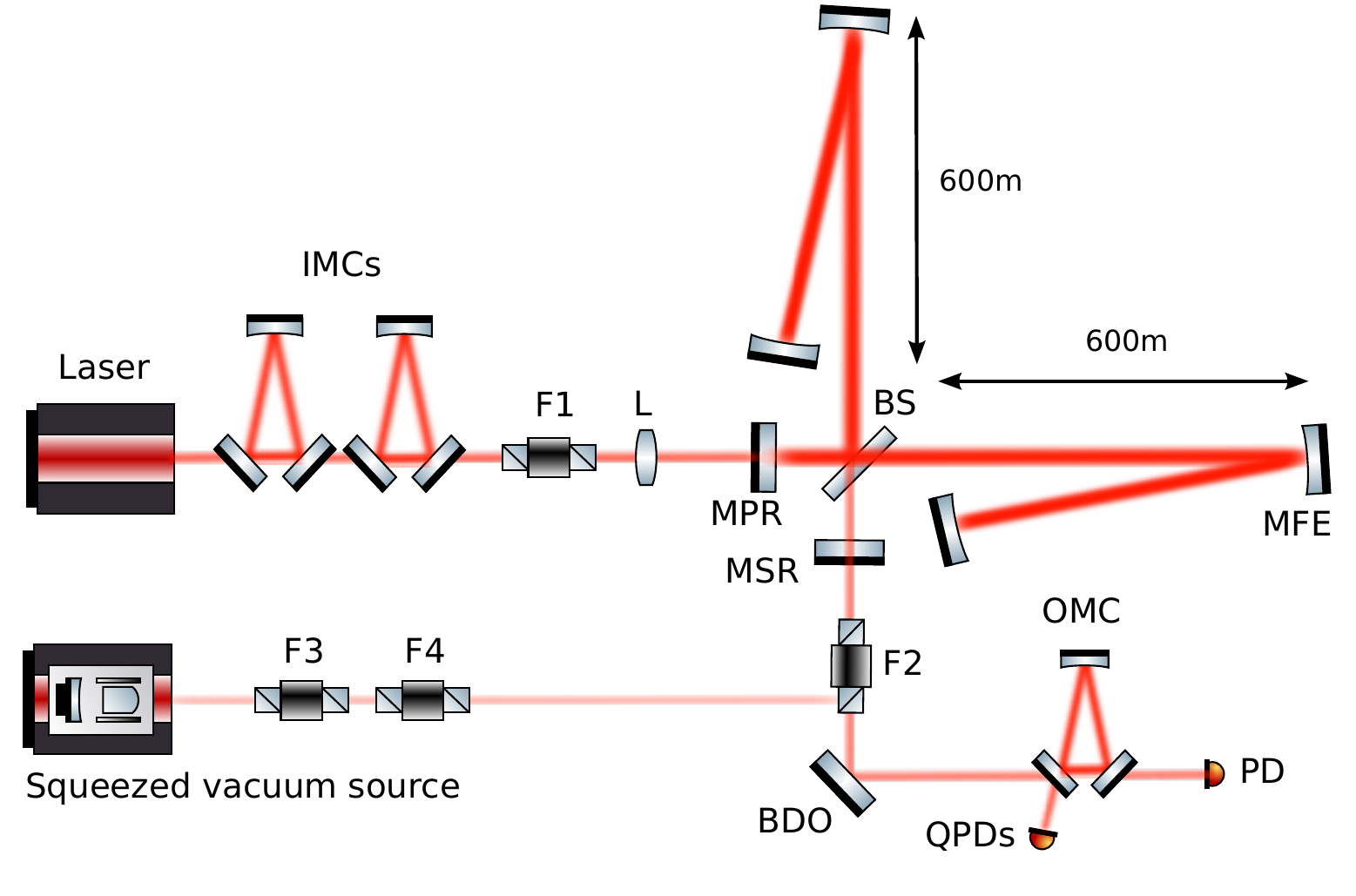}
\caption{Simplified optical layout of GEO\,600. IMCs: input mode cleaners,
F1, F2, F3, F4: Faraday isolators, L: lens, 
MPR: power recycling mirror, MSR: signal
recycling mirror, BS: beam splitter, MFE: east end mirror, 
BDO: output beam director, OMC: output mode cleaner, 
PD: photodiode for DC readout, QPDs: quadrant photodiode for 
squeezed field and OMC alignment.}
\label{fig:layout}
\end{figure}

GEO\,600 is the German-British laser-interferometric gravitational
wave (GW) detector with 1200\,m long arms folded inside 600\,m long
beam tubes located near Hannover, Germany \cite{Willke2002GEO,
GeoStatus2008, Grote2010GEO, Dooley2015GEOStatus}. 
Figure~\ref{fig:layout} shows a simplified optical layout of GEO\,600,
referred to throughout this paper.

The GEO\,600
detector stood apart from its contemporary first-generation
counterparts because of a significantly limited budget and the incorporation 
of more innovative but also
riskier technologies \cite{Affeldt2014Advanced}. After
conducting science runs together with the other first-generation GW
detectors, the GEO\,600 detector continued to operate as a data-taking
instrument with the LIGO Hanford 2\,km interferometer in a program called
Astrowatch from 2007 to 2009. 
During this time the other detectors in the network, LIGO and Virgo, 
were enhancing their 4\,km and 3\,km instruments,
respectively. \cite{JoshSmithEnhancedAdvanced, Acernese2008Virgo}. Then, in
2009, an upgrade program called GEO-HF began with the goal of carrying out a
series of upgrades to improve the detector's high-frequency (HF) 
shot-noise-limited sensitivity, 
around 1\,kHz and above,
and to demonstrate new technologies~\cite{Luck2010Upgrade}. 
Simultaneously, GEO\,600 was to serve as the
GW community's Astrowatch detector, continuing to collect scientific
data while the other detectors went offline for a 4--6 year period to
construct and commission the so-called Advanced
Detectors~\cite{TheLIGOScientificCollaboration2015Advanced,
  Acernese2015Advanced}. GEO\,600 thus operated as an observatory for
large parts of the GEO-HF program, mandating substantial effort to
maintain continuous operation. Not only did such effort facilitate the
small chance of serendipitous discovery, but it also provided a unique
opportunity to test new technologies in an observatory-style
environment, setting GEO\,600 apart from a standard laboratory
setting.

This article explains the detector developments during the course of
the GEO-HF program, reviewing the early upgrades that were reported in
the latest update article \cite{Grote2010GEO} and are continuing to the
present day. Section \ref{sec:geohf} begins with a review of the
GEO-HF goals including the design noise curve and provides a summary
of the accomplishments to date. Each of the upgrades is then discussed
in more detail, including how they help accomplish the GEO-HF goals
and the successes and challenges presented by each. Section
\ref{sec:challenges} summarizes the overall current state of the
detector, highlighting an up-to-date detector noise budget and
challenges faced in moving forward. The potential effectiveness
of GEO\,600 as a GW detector and the tools developed to aid its operation as an
observatory while carrying out installation and optimization work in parallel (called \emph{commissioning}) are described in Section~\ref{sec:observatory}. Finally,
Section~\ref{sec:summary} provides a summary and outlook.


\section{GEO-HF Upgrades}
\label{sec:geohf}

\begin{figure}[tb]
\centering
\includegraphics[width=1.0\columnwidth]{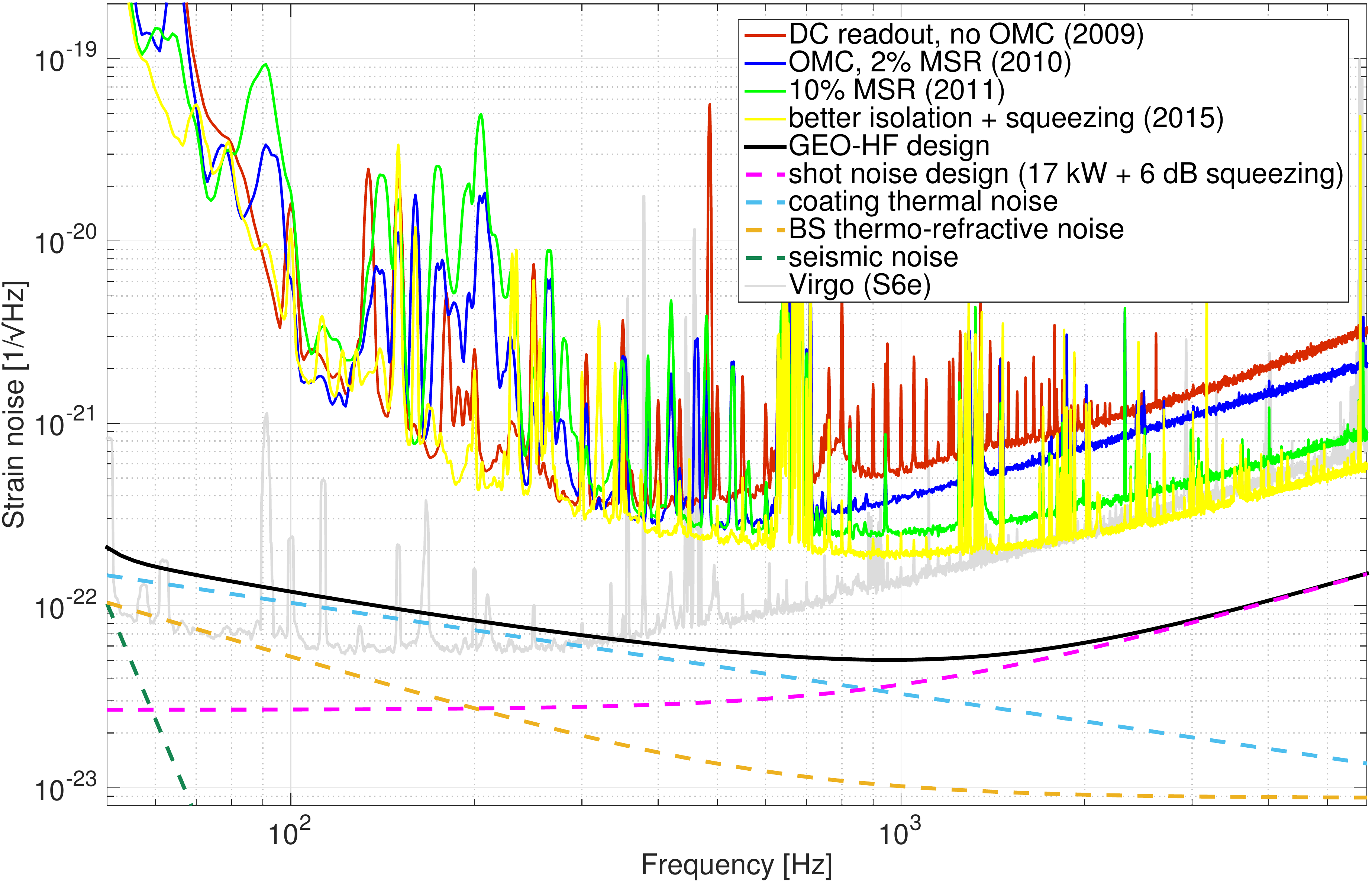}
\caption{Progression of strain-equivalent noise curves upon completion of
major milestones of the GEO-HF upgrade program compared to
the GEO-HF design noise curve. A new
  signal recycling mirror (10\,\% MSR), 
  squeezed vacuum injection, and DC readout
  contribute to the substantial improvement in high-frequency
  sensitivity. The change to tuned signal recycling with DC readout
  was made early on in the GEO-HF program and is reflected in all of
  these curves. An increase in laser power is still a work in progress. 
  The GEO-HF design noise is dominated by shot noise and coating thermal noise.
  Shot noise is plotted for the GEO-HF goals of 20\,W 
  input power (17\,kW circulating) and 6\,dB squeezing. 
  (See Figure~\ref{fig:NB} for a comparison of strain noise to its corresponding shot noise.)
  The coating thermal noise \cite{Harry2012} is dominated by the 
  (approximately equal) contributions from the four mirrors of the long arms.  
  Thermo-refractive noise of the beam splitter (BS) \cite{Benthem2009} 
  is significant for GEO\,600 due to the lack of arm cavities. The large cluster of lines centered around 650\,Hz are the violin modes of the
  suspension fibers.
  A strain noise curve for Virgo during the S6e science run is shown 
  in light grey for comparison~\cite{VirgoVSR4}.
}
\label{fig:h}
\end{figure}

The limiting fundamental noise sources for the GEO-HF design are
plotted in Figure~\ref{fig:h} and are compared to strain sensitivity
curves highlighting selected upgrade milestones.  The GEO-HF goal is to
reach a sensitivity limited by fundamental shot noise and to maintain or improve low
frequency sensitivity. The parameters that define the shot noise
design curve are 20\,W input power (17\,kW circulating power), 6\,dB of
observed squeezing (defined as the improvement of the shot-noise-limited
sensitivity due to the application of squeezing), and tuned signal recycling 
with a signal recycling mirror of 90\,\% reflectivity. The following incremental
upgrades to the baseline GEO\,600 configuration are the cornerstones
of the GEO-HF program:
\begin{itemize}
\item{a transition from RF readout to DC readout, and the addition of an
  output mode cleaner (OMC).}
\item{a change from detuned to tuned signal
  recycling and a replacement of the signal recycling mirror to one with
  higher transmission.}
\item{the injection of squeezed vacuum into the detection port.} 
\item{an increase in laser power and measures supporting the operation
  at higher power.}
\end{itemize}

To date, each of these upgrades has been partially or fully
implemented and as much as a factor of 4 improvement in sensitivity
above 600\,Hz has been achieved. In the sections below we review the
motivations \cite{Luck2010Upgrade} and present the implementation,
results, and implications of each of the upgrades. The discussions are
grouped by topic and do not necessarily reflect chronological
order. Refer to Figure \ref{fig:timeline} for a timeline of major
events during the GEO-HF program.


\subsection{DC readout and implementation of an OMC}
\label{sec:DCOMC}

Amongst the first of the upgrades in the GEO-HF program was the change
of the GW readout scheme from RF (heterodyne) to DC (homodyne) readout
in 2009 \cite{Hild2009DCreadout}. DC readout provides a fundamental
$\sqrt{3/2}$ improvement in sensitivity at shot-noise-limited
frequencies due to the elimination of the cyclostationary shot noise
of RF readout \cite{GeaBanacloche1987Squeezed, Niebauer1991Nonstationary}. 
It was also an important simplification for the implementation of squeezing because a
squeezed light source needs to be prepared only in the audio band, rather than
at both, audio and RF, frequencies. Moreover, DC readout significantly
reduces the coupling of oscillator phase and amplitude noise to the GW
signal \cite{Fricke2012DC}. Phase noise of the electronic RF oscillator
and of the RF signal path had been a nearly
limiting noise source for GEO\,600~\cite{Grote2010GEO}
with RF readout. 
With DC readout, the phase noise coupling to the GW readout 
was reduced by a factor of 50. 
DC readout also has disadvantages.
Due to the presence of the
TEM$_{00}$ carrier field created by the dark fringe offset,
alignment sensing errors for the main interferometer 
can be induced by beam motion on the alignment sensors.
Stable locking with DC readout was possible only after a technique
known as 2f-centering was developed, which actively
stabilizes the beam on the alignment sensors, as further
described in \cite{Grote2010GEO}.

The introduction of an output mode cleaner (OMC in Figure~\ref{fig:layout}) 
followed shortly thereafter in 2010 \cite{Prijatelj2012Output}. A four-mirror
monolithic bow-tie cavity was installed in the interferometer's output
beam path in a new vacuum tank separated by a window from the rest of the
vacuum system for ease of access. An OMC, whether in conjunction with
RF or DC readout, is necessary in order to achieve the best possible
classical shot noise limit. The OMC suppresses higher-order 
modes (HOMs) of the optical light field which originate from contrast defects within the main interferometer. 
When used with DC readout, an OMC
also attenuates the RF modulation sidebands used to control auxiliary interferometer degrees of freedom. If the output field is not
filtered, both higher-order modes and RF sidebands increase shot noise
without contributing to the GW signal.

The stark effect that the OMC had on reducing GEO\,600's high
frequency noise floor can be seen in Figure~\ref{fig:h}. The trace
labeled `OMC, 2\% MSR (2010)' can be compared to the trace labeled `DC
readout, no OMC'. Both spectra were taken with the detector in tuned
signal recycling mode with DC readout. Although the use of an OMC
indeed improved high-frequency sensitivity, it initially degraded the
GEO\,600 sensitivity at lower frequencies. The root of the problem
comes from the fact that higher-order modes in the frame of the OMC can
couple into the GW signal via jitter of the output beam. 
The amplitude of the coupling is non-stationary and the spectral features around 90\,Hz and 200\,Hz and at nearby
frequencies in trace `OMC, 2\% MSR' in Figure~\ref{fig:h} were caused
by this process.


Significant improvements in the low-frequency sensitivity achieved since the
introduction of the OMC have come as a result of addressing the amount of
jitter motion and its coupling to the readout. This includes an upgrade of the
output optics'
(BDO in Figure~\ref{fig:layout})
suspensions to reduce beam jitter and an improvement of the vibration
isolation of the OMC.
Furthermore a 37\% reduction in HOMs at the output port was achieved
through thermal compensation of an astigmatism at the east end mirror
(MFE in Figure~\ref{fig:layout}),
thus potentially decreasing coupling of HOMs to the GW channel as well
\cite{Wittel2014Thermal}.  The effects of these efforts is seen by
comparing the low-frequency sensitivity of the curve in
Figure~\ref{fig:h} labeled `better isolation' with those preceding it.

The development of an OMC alignment scheme in the presence of higher
order modes and beam jitter is also critical for reducing the coupling
of beam jitter and HOMs to the GW readout. Originally, a simple dither
of the output optics' suspensions was used in a control scheme to maximize the power throughput of the OMC. However, because 80\,\% of the power at the GEO\,600
output port is composed of HOMs, such a scheme may find an alignment
that partially transmits HOMs rather than pure TEM$_{00}$ light. A
scheme that maximizes only the TEM$_{00}$ light transmitted through
the OMC came next by adding an audio frequency modulation to the
carrier field \cite{SmithLefebvre2011Optimal}. 
The response of the GW readout (the optical gain) is maximized
with such a double-demodulation scheme, but
the signal-to-noise ratio (SNR) of these alignment signals is low and 
creates a significant limitation
on the alignment control bandwidth. The newest technique, modulated
differential wavefront sensing (MDWS), uses wavefront sensors in
reflection of the OMC (QPDs in Figure~\ref{fig:layout}) and has been recently commissioned
\cite{Affeldt2014Advanced}. The wavefront sensors sense the relative
alignment between the interferometer RF sidebands representing the GW
mode and audio sidebands generated from a length dither of the OMC
representing the OMC eigenmode. This scheme both eliminates low
frequency dithering of the output steering optics and facilitates an
increase in the alignment control bandwidth up to several Hz, helping to reduce the non-stationary beam jitter coupling.




\subsection{Signal recycling bandwidth increase}

Other relatively early upgrades which enabled progress towards reaching
the GEO-HF goals were related to the signal recycling cavity (SRC). 
An important step in the GEO-HF program was the reshaping of the shot noise 
by changing the finesse of the SRC and varying its operating point (detuning).

Prior to the GEO-HF upgrade, the signal recycling mirror 
(MSR in Figure~\ref{fig:layout})
was held off-resonance from the carrier in order to maximally amplify one of
the GW sidebands. In this state, known as detuned signal recycling, an
increase in the GW signal-to-shot noise ratio is achieved in a band
around a particular frequency at the expense of all other
frequencies. From 2005 to 2009, GEO\,600 used a detuning of 530\,Hz.
Early experiments showed that operating the signal recycling cavity in
the detuned state was not optimal for GEO\,600 due to technical noise
couplings to the GW readout which could be significantly reduced in a
tuned state where upper and lower GW sidebands are equally resonant
\cite{Hild2007Demonstration}. Operation in a tuned state would not
only improve the high-frequency sensitivity, but also open the path to
being able to benefit from squeezing with a frequency-independent quadrature angle at all
shot-noise-limited frequencies.  In 2009, tuned operation of the signal recycling cavity was commissioned together
with the change to DC readout \cite{Grote2010GEO}. All curves in Figure~\ref{fig:h} use tuned signal
recycling with DC readout. A comparison of the tuned and detuned cases
is found in Ref.~\cite{Hild2007Demonstration}.


The use of tuned signal recycling is most appropriate if the bandwidth
of the SRC is comparatively high. The build-up of the GW signal falls
off as a Lorentzian with frequency, so a large improvement in high
frequency sensitivity could therefore be achieved by decreasing the
finesse of the SRC. This was accomplished in fall 2010 by exchanging
the signal recycling mirror from one with a reflectivity of $R=98\%$
to one with a reflectivity of $R=90\%$. The cavity bandwidth increased from about 230\,Hz to
1150\,Hz. The trace labeled `10\% MSR' in Figure~\ref{fig:h} shows the
result of the signal recycling mirror swap in comparison to the `2\%
MSR' trace. An improvement by a factor of more than 2 at high
frequencies was achieved as planned.

The new signal recycling cavity's lower finesse had a number of
negative side effects, though, which resulted in sensitivity setbacks at lower
frequencies. The exaggeration of the peaks from 80\,Hz to 1\,kHz in the
`10\% MSR' trace in Figure~\ref{fig:h} compared to before
results largely from a reduction of the phenomenon known as mode
healing \cite{Strain1991Experimental}. With the lower MSR reflectivity, HOMs at the
interferometer's output port are less effectively converted
back into the interferometer's fundamental mode. The HOM content at the output port
increased by about 50\%. Given that HOMs are the means by which beam jitter
couples to the GW readout and that beam jitter had not yet been
reduced at the time of these spectra, the dominant reason for an
increase in low-frequency noise was a higher coupling of jitter
through the OMC.


Another side effect of the new signal recycling cavity's lower finesse
was that it decreased the SNR of the SRC length signal, which is
derived from 9 MHz resonant sidebands and the carrier light in the
SRC. The result of a lower SNR signal while maintaining the same
control bandwidth (35\,Hz) is higher feedback noise. This is currently
the limiting noise source from 50\,Hz to 100\,Hz, as will be discussed
in Section~\ref{sec:challenges}.


\subsection{Squeezing}

\begin{figure}[tb]
\centering
\includegraphics[width=0.8\columnwidth]{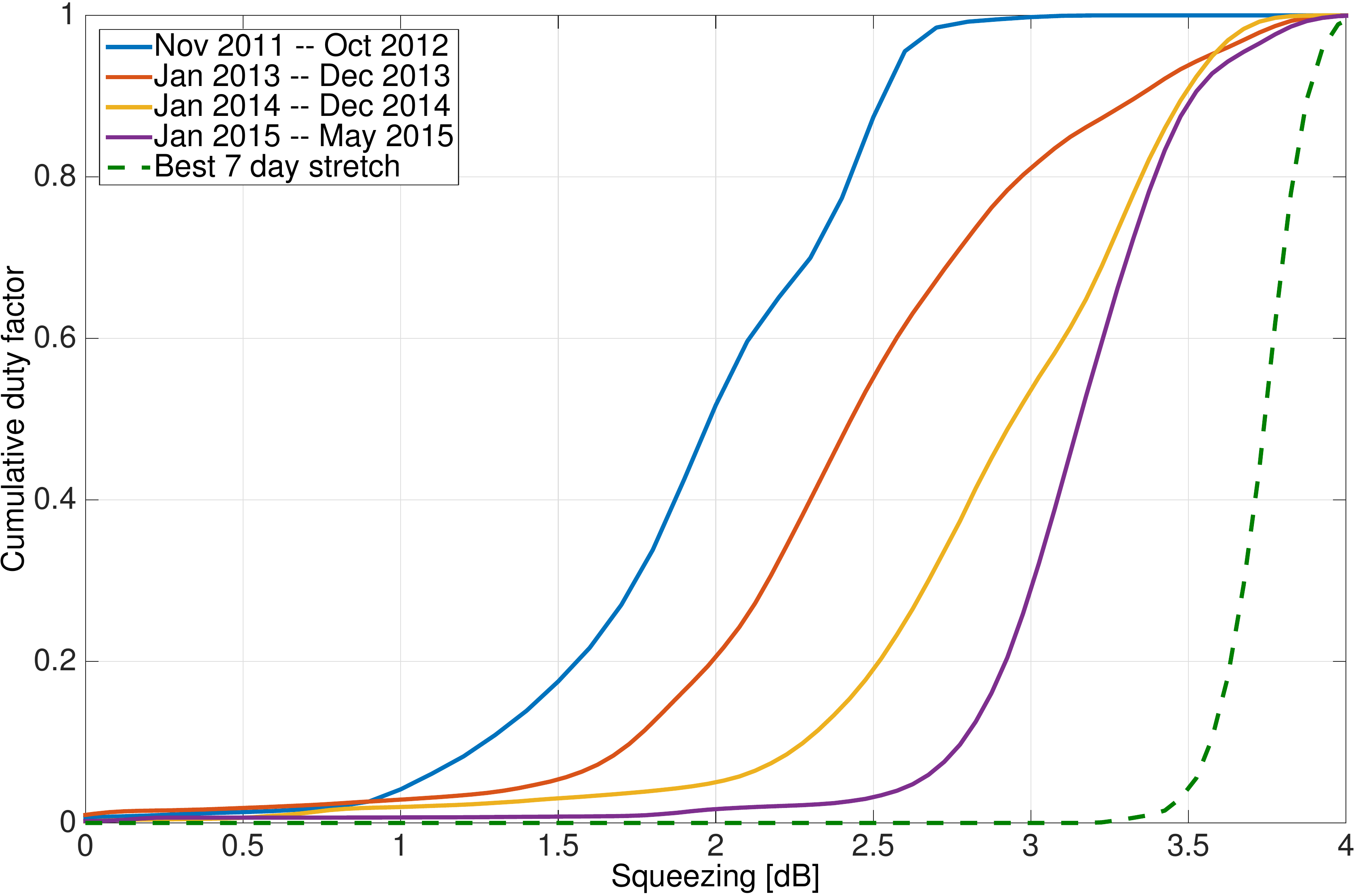}
\caption{Record of the improvement of the observed squeezing level
  during the course of the last three and a half years. 
  The cumulative duty factor while squeezing was applied illustrates for what fraction
  of time squeezing at a given level (and lower) was observed. 
  (A cumulative duty factor of 0.5 corresponds to the median of the distribution.)
  Four consecutive periods of time are displayed. 
  A 7 day example of a particularly stationary period is
  shown in order to demonstrate what can be achieved on shorter time
  scales. During all Astrowatch science times during this period, the
  squeezing up time was around 85\,\%.}
\label{fig:sqz}
\end{figure}

Injection of squeezed vacuum states to the interferometer's output port is a
novel technique employed to suppress quantum noise
\cite{Caves1980QuantumMechanical, Schnabel2010Quantum} and is one of the technology
demonstration highlights of the GEO-HF program. The GEO\,600 squeezer
was built at the Albert-Einstein-Institut in Hannover and transported to the site in April 2010
\cite{Vahlbruch2010GEO, Khalaidovski2011Beyond}. Following
installation and integration, a first successful demonstration of the
high-frequency noise reduction was achieved in fall 2010 just prior to
the signal recycling mirror swap. The sensitivity above approximately
1\,kHz was improved by 3.5\,dB and this
accomplishment marked the first of its kind for large scale GW
detectors \cite{2011Gravitational}. A typical example of how squeezing
helps achieve the GEO-HF goals can be seen by comparing the traces
labeled `squeezing' in Figure \ref{fig:h} to `10\% MSR' at frequencies
above approximately 700\,Hz.

The sensitivity gain from
squeezing at GEO\,600 is obtained without negative side effects, although data quality was initially compromised upon the squeezer's installation and some indirect noise contamination required mitigation efforts.
The main challenges encountered were from the back-scattering of light into the interferometer mode from the additional optics required for squeezing and noise from control malfunctions of the squeezer.  The
scattering is mitigated through acoustic and vibrational isolation of
the optics, and three Faraday isolators in the squeezing path including the injection Faraday, each of
which provide approximately 40\,dB of isolation (F2, F3, F4 in Figure~\ref{fig:layout}). 
This isolation is sufficient, but any reduction cannot be afforded. Additional
controls implemented on optical and mechanical degrees of freedom both
internal to the squeezer and on those relative to the main
interferometer, in addition to other software enhancements, allow the
squeezer to operate routinely and consistently without the
introduction of spurious noises into the detector.
An analysis of the data glitchiness with and without squeezing was
performed in 2012 and shows that squeezing does not negatively affect
the data quality \cite{Grote2013First}.


Throughout the work on squeezing, the goals have been to ensure stable
operation and to reach a high stable squeezing factor.  
The resulting progress is shown in Figure~\ref{fig:sqz}. 
The cumulative duty factor with squeezing is plotted for four 
consecutive periods of time. Squeezing has been in near continuous use
since fall of 2011 and in all has operated for 85\% of the time during 
which the detector was in science mode.
The average observed squeezing level has progressed by several
tenths of a dB per year. 

The development and commissioning of squeezer alignment and phase
control systems has had top research priority and contributed to the
improvements in Figure~\ref{fig:sqz}. A wavefront sensing alignment
scheme in reflection of the OMC was designed to ensure maximum overlap
of the squeezed vacuum field with the interferometer output field
\cite{Schreiber2014Alignment}. It operates with a bandwidth of up to
several Hertz and is important for long-term stability of the squeezing
level.
In addition, a new error signal derived in transmission of the OMC
was devised for sensing the relative phase of the squeezed field with
the interferometer output. It eliminates lock point offsets due to
HOMs that had previously created a fluctuating squeezing level
\cite{Dooley2015Phase}. Both the new alignment and phase control
schemes have been commissioned and are in continuous use.

A new focused research effort is now underway to reduce the optical losses.
The optical losses in the GEO\,600 squeezer path amount to approximately 40\%, and are the dominant limitation on the potential sensitivity improvement from squeezing. The losses alone reduce the maximum possible observed
squeezing from 10 dB to 4 dB. The poor optical efficiency results quite easily from the sum of many small losses, including the 8 passes through polarizing beam splitters, OMC internal losses, imperfect mode matching to the OMC, and photodetector quantum efficiency. Carefully addressing each of these loss sources to reduce the total losses to 20\% is a near-term goal. Phase noise also contributes to reducing the maximum possible squeezing level. It has been measured to be about 37\,mrad rms and reduces GEO\,600's best observed squeezing by about 0.2\,dB. The record to date is an improvement of the
shot-noise-limited strain sensitivity by 3.7~dB, the equivalent to a laser power increase by a factor 2.3~\cite{Grote2013First}.

\subsection{Power increase}
\label{sec:power}

\begin{figure}[tb]
\centering
\includegraphics[width=0.8\columnwidth]{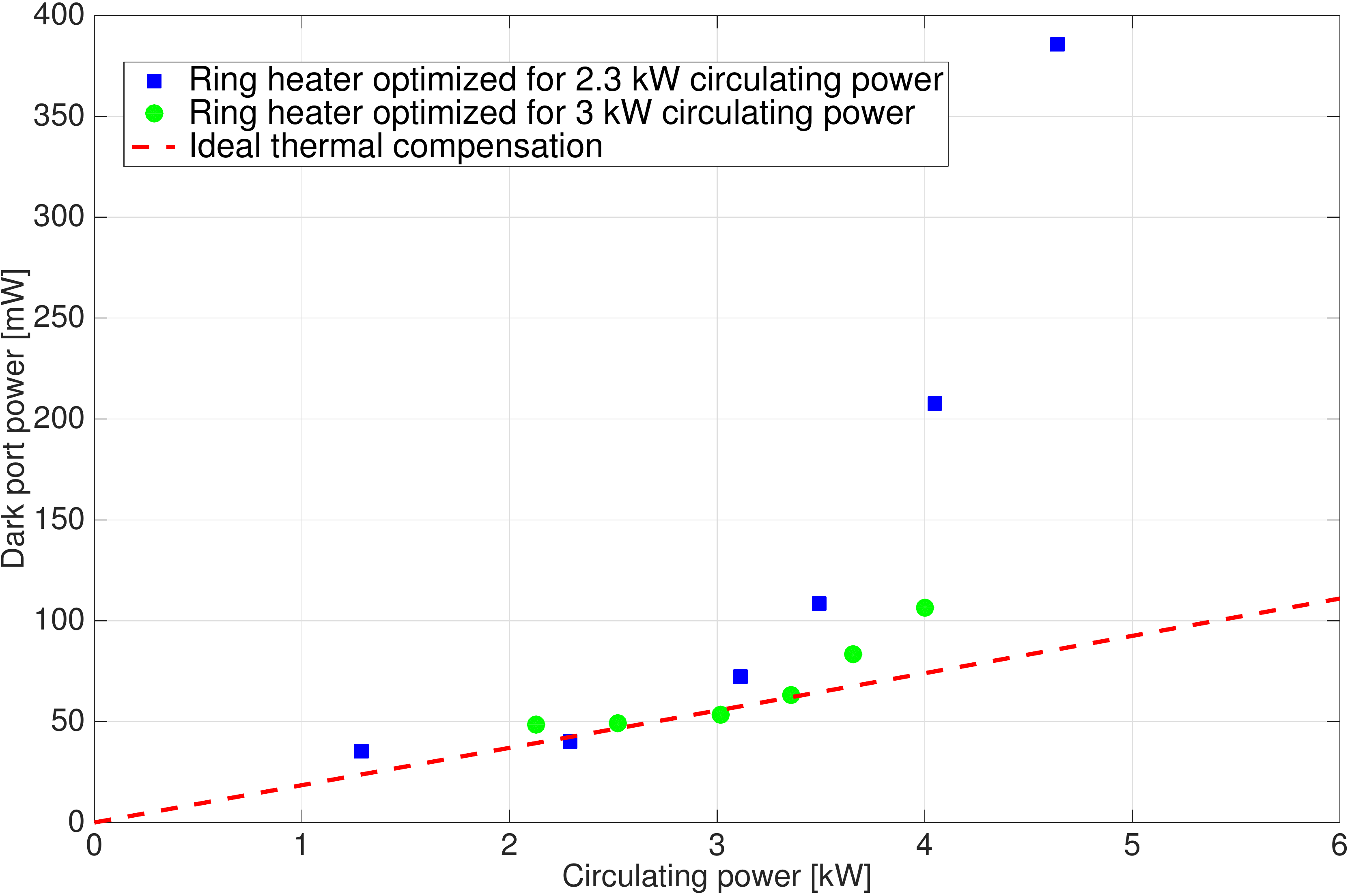}
\caption{Measurement data demonstrating the fundamental reason for the difficulty
  in increasing the laser power in GEO\,600. A power-dependent thermal
  lens in the BS results in an increase of higher-order modes at the
  output port which in turn leads to a breakdown of interferometer
  control. Indirect compensation of the BS thermal lens through a
  radius of curvature adjustment of the east arm's folding mirror offers some
  relief. The dashed red line
  shows the expected dark port power if the BS thermal lens could be
  fully compensated. A constant contrast defect of 18 ppm is assumed.}
\label{fig:thermallens}
\end{figure}

The planned power increase has proven to be the hardest of the GEO-HF goals to meet.
An increase from 2\,kW circulating power to approximately 17\,kW (20\,W input power) should
provide a factor of about 3 improvement in sensitivity at frequencies limited by shot noise (which should be 900\,Hz and
above, in the absence of technical noise sources). To date, stable operation
of the detector with up to 4\,kW circulating
power has been achieved and the corresponding expected high-frequency
sensitivity improvement observed. However, due to increasing noise of unknown
origin at frequencies below about 1\, kHz, this higher power state is not yet
used in standard running conditions as is discussed in Section~\ref{sec:challenges}.

A number of hardware upgrades had to be carried out to ensure the
availability of at least 17\,kW circulating power. These upgrades began in
fall 2011 with the installation of a 35\,W laser system, 
to replace the previous 12\,W laser. 
The new system consists of a laser-diode-pumped Nd:YAG master laser 
and a subsequent Nd:$\rm YVO_4$ amplifier stage
\cite{Frede2007Fundamental, Kwee2012Stabilized}.
The following fall, a vacuum incursion took
place to replace several components of the input optics chain to
ensure high-power compatibility. The LiNbO$_3$ crystals used in the
electro-optic modulators (EOMs) were replaced with RTP crystals that
have more than twice the damage threshold and a lower thermal lensing
effect \cite{Bogan2015Novel}. Four of the six input mode cleaner mirrors 
were replaced to
reduce the finesse of the cavities in order to improve their
transmission and ease the locking. 
In addition, the in-vacuum mode matching telescope was adjusted to 
preemptively compensate for thermal
lensing that would result from power absorption at high power in the
input chain Faraday isolators and EOMs \cite{Affeldt2014PhD}.
An interchangeable lens 
(L in Figure~\ref{fig:layout}) on a suspended bench was replaced after simulations showed that the previous choice of 
lens was not optimal for high-power operation. The thermal lens effect
to be expected from the Faraday isolator (F1) was measured in the laboratory,
and served as input for the optical simulations. 

A challenge in achieving higher
power operation which was addressed early on in the GEO-HF program was
that the signals on the individual optics' local position sensors were
contaminated by scattered light from the main interferometer. 
The amount of scattered light increases proportional to the circulating 
power in the main interferometer optical mode and would thus inject 
more and more false position information to the local
velocity damping servos as the power would increase.
As a result, the interferometer became
more unstable for higher circulating power and would eventually lose lock. 
The elimination of stray light coupling was achieved through the
implementation of a modulation-demodulation technique for the local
position sensors.
The currents of the light emitting diodes of the 
sensors are modulated at single frequencies
around 7\,kHz--10\,kHz. The photocurrents of the adjacent photodiodes
are coherently demodulated and this signal serves as the error 
signal for the velocity damping servo \cite{Affeldt2014PhD}.

The single most critical limitation to operating GEO\,600 with higher
laser power is a strong thermal lens at the beam splitter (BS).
Because GEO\,600 does not have Fabry-Perot arm cavities, all of the
circulating power in the interferometer is transmitted through the
substrate of the BS, whose absorption has been measured to be
0.5\,ppm/cm \cite{Affeldt2014Advanced}. Furthermore, due to the beam's non-normal angle of incidence on the BS, the resulting lens is elliptical and more challenging to compensate. The severity of the
uncompensated thermal lens is highlighted in
Figure~\ref{fig:thermallens} which shows measurements of the quadratic increase of
power at the output port as a function of power at
the BS due to contrast defect. In the case of a perfectly compensated thermal lens, the
relationship should be linear.

Partial, indirect compensation of the BS thermal lens is accomplished
through the use of a radiative ring heater behind the far
east mirror which had been installed in 2003 to correct for
the mirror's slightly incorrectly manufactured radius of
curvature~\cite{Luck2004Thermal}. The data points in Figure \ref{fig:thermallens} show that optimal ring heater powers that compensate the BS thermal lens can be found for up to 3\,kW circulating power. Exploration of optimal ring heater powers for higher circulating power levels has been challenging because of the time required for thermal equilibrium to be reached and because of an increasing lack of detector stability. For circulating powers beyond about 3.5\,kW, the excess output port power
increases the shot noise on the many auxiliary sensing photodiodes, and
contributes to a break-down of control and loss of lock.

A next step in the series of work required to achieve higher laser
power for GEO-HF is to install a thermal compensation system directly
at the BS. An array of heating elements has been designed to project optimized patterns
of radiation onto the BS. Initial
results have demonstrated a 30\,\% reduction of thermally induced HOMs
and commissioning is ongoing. 



\section{Noise budget and challenges}
\label{sec:challenges}

\begin{figure}[tb]
\centering
\includegraphics[width=\columnwidth]{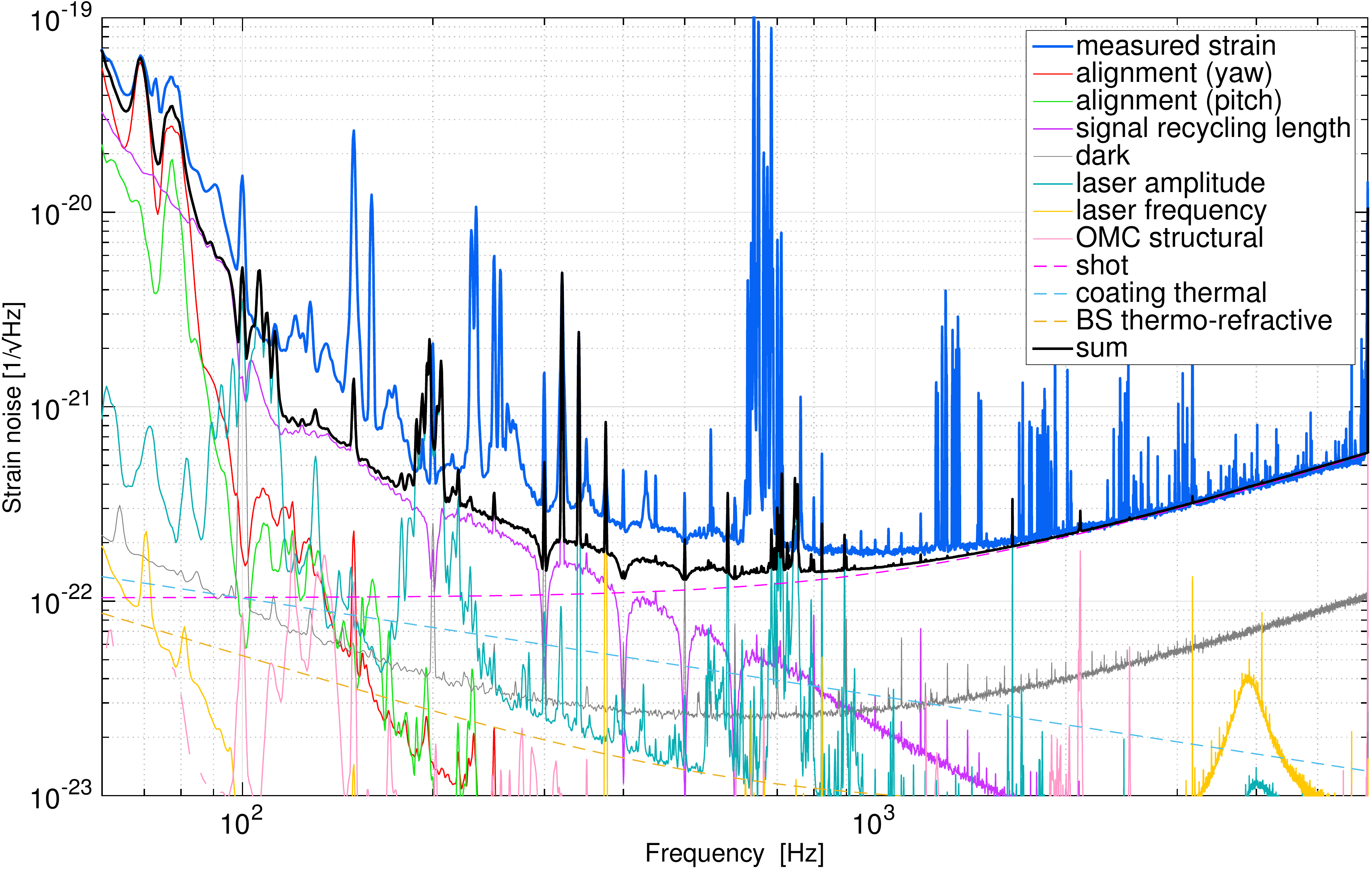}
\caption{Noise budget of GEO\,600 with 2\,kW circulating power and 2.9\,dB
  observed squeezing. The blue curve is the measured amplitude spectral
  density of the strain-equivalent noise (also called sensitivity throughout
  the paper). The dashed lines represent analytical models
  of fundamental noise sources and the solid lines are actual
  data. The uncorrelated sum of fundamental and technical noise sources
  explains the GEO\,600 sensitivity above 1.5\,kHz and partially below 100\,Hz. 
  Excess noise in the mid-frequency band remains unexplained.}
\label{fig:NB}
\end{figure}

Figure~\ref{fig:NB} shows the current status of the understanding of
GEO\,600's technical and fundamental noises. This snapshot was taken
during a night of standard Astrowatch operation with 2\,kW
circulating power and 3.2\,dB of observed squeezing. The dashed lines
are the analytical models of fundamental noise sources as presented in
Fig.~\ref{fig:h} and the solid lines are actual data. The noise projections are created from continuously measured data and recorded coupling transfer functions \cite{Smith2006Linear}. To aid commissioning, the noise budget is
updated in near real time once every few seconds. The transfer
functions that are used to make the projection are updated
periodically. The laser
frequency stabilization control loop and the SRC length control loop have calibration lines to enable
real-time tracking and adjustment of the magnitude of the
corresponding noise coupling transfer functions.  
The uncorrelated sum of the fundamental and technical noise sources can be compared to the
measured strain sensitivity and shows that the GEO\,600 noise floor is well
understood above 1.5\,kHz and mostly understood below 100\,Hz. However, between 100\,Hz and 1.5\,kHz there is significant unexplained noise.

Below 100\,Hz the measured noise often matches
the projected noise. In this frequency region the strain sensitivity is dominated 
by technical control noises. 
One reason for the control noises being high is that there is
only very limited seismic pre-isolation. The control loops therefore
need comparatively high bandwidth to keep the interferometer
sufficiently quiet. The alignment control loops and the signal recycling cavity length control loop are
the most severe sources of noise impression below 100\,Hz. The
wavefront sensors that derive the Michelson alignment signals are at the output port of the interferometer and
have a noise floor at high frequencies dominated by shot noise from higher
order modes. Below approximately 100\,Hz, these signals are limited by
beam jitter noise coupling on the associated wavefront sensors.  Unity
gain frequencies of about 5\,Hz are necessary for these loops to have
enough low-frequency gain to control the pitch and yaw modes of the suspensions. The
SRC length loop has a bandwidth of 35\,Hz such that (shot-noise
limited) feedback noise is a factor of a few below the strain noise
floor.
Partial cancellation of this noise with a feed-forward scheme proved
difficult, as the level of the signal recycling mirror length noise
coupling to strain varies with output beam alignment onto the OMC.
The process is the same as for the beam jitter coupling in conjunction
with higher order optical modes, as described in Section
\ref{sec:DCOMC}.

The strain sensitivity at high frequencies is dominated by shot
noise. The next most significant contributor, dark noise from the
readout electronics, is a factor of 7 below shot noise without
squeezing, but is only a factor of 4.5 below shot noise with squeezing
engaged, as shown in Figure~\ref{fig:NB}. Work is ongoing in
implementing a new electronics design to decrease dark noise by a
factor of 2 to 3 in order to increase the level of observable
squeezing. Laser amplitude and frequency noise are not always
stationary and they sometimes start to affect the sensitivity at several
kHz, as can be seen in this example from the laser amplitude noise
peak at 4\,kHz~\footnote{The cause for this non-stationarity is under
investigation.}.

In the search for the culprit of the excess mid-frequency noise, many
potential sources of technical noise have been ruled out. This
includes saturations at both RF and DC of sensing
photodiodes. Scattered light has been systematically searched for
through extended tapping tests for acoustic coupling via acoustic
excitation of vacuum chambers and associated optics. 
Filter experiments 
have ruled out scattering from sources external to the
vacuum system. Suspended baffles were installed around the east 
arm's end test mass and the two main test masses in the central area
to rule out small-angle scattering. 
Efforts to think
about potential additional fundamental noise sources has in fact
resulted in an updated thermal noise model for GEO\,600 because of a
stripe pattern (caused by the folded arm geometry of GEO\,600) 
on the far end mirrors that had not been previously
considered \cite{Heinert2014Thermal}. As a result of this
investigation, the theoretical coating thermal noise curve has been
adjusted to be 20\% higher than in the last update article, but is
still too far below the noise floor to explain the observed
strain-equivalent noise. Furthermore, the amount of excess noise is not stationary and studies to determine the time scale of its changing amplitude produce varying results. 

As mentioned in Section \ref{sec:power}, operation with 4\,kW circulating power can be achieved during commissioning
periods. Operation with 2\,kW to 2.5\,kW is standard, however, because the
increase in laser power brings with it an increase in noise below
600\,Hz. The shape of the excess noise is $1/f$ and despite systematic
investigations to identify its cause, this problem has not yet been
solved. An increase in laser power remains the last of the GEO-HF
goals to be carried out and hinges on successful implementation of
direct thermal compensation at the beam splitter and uncovering the
source of the power-dependent mid-frequency noise.


\section{GEO\,600 as a GW detector}
\label{sec:observatory}

An integral aspect of the GEO-HF program was not only the series of
upgrades to improve the sensitivity at high frequency, but also GEO\,600's status as a GW detector. The
standard mode of operation during the last 6 years has been for
commissioning activities to take place during the day and for GEO\,600
to operate in Astrowatch mode during the nights and weekends. An
exception was the period from June to August 2011 when GEO\,600
participated in a dedicated data-taking run (known as S6e or VSR 4)
with the French-Italian detector, Virgo. At this time, the GEO-HF
upgrades that had already taken place resulted in GEO\,600's
sensitivity above 2\,kHz equalling that of the Virgo detector, as can be seen
in Figure~\ref{fig:h} (light grey curve) \cite{LIGOScientificCollaborationandVirgoCollaboration2014Methods}.

\begin{figure}[tb]
\centering
\includegraphics[width=0.8\columnwidth]{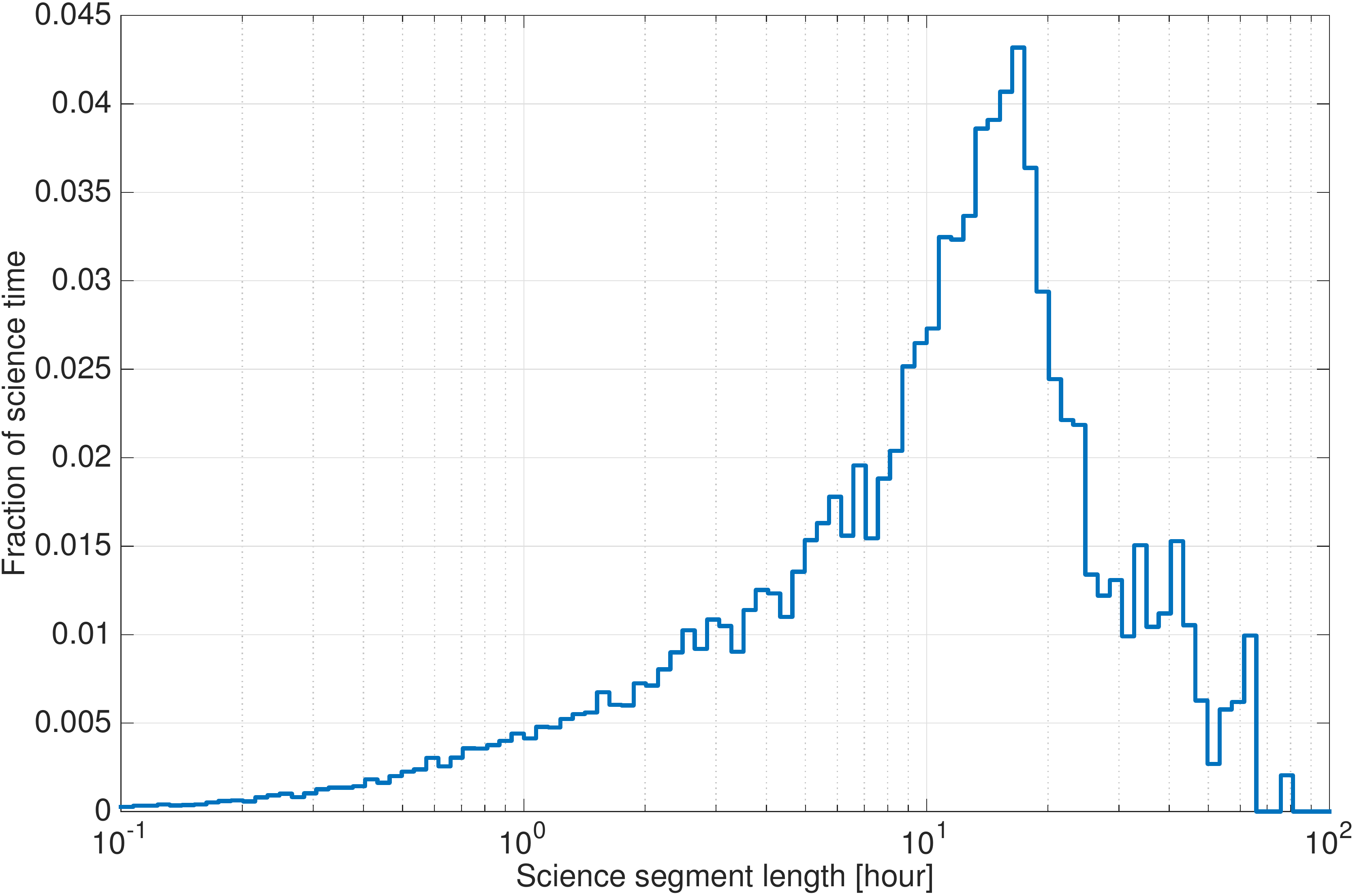}
\caption{Histogram of the fraction of science time spent in a lock of a
  given length during Astrowatch times from 2008 through 2014. The total
  duration is 38400 hours or 4.4 years, resulting in a 63\,\% average science time duty factor.}
\label{fig:locks}
\end{figure}

In all, during the 7 years between 2008 and 2015, 63\% of the time was spent collecting science quality data. The achievement of this high duty factor is possible because of the detector reliability which results in long continuous lock stretches and because of the success of automatic re-locking which requires human intervention on average only once every five days.
Figure \ref{fig:locks} shows
a histogram of lock stretch lengths during the course of
the GEO-HF upgrade program. The median segment duration is 11.5 hours and the longest lock stretch to date occurred in early January 2015 and
was 102.5 hours long. Despite this reliability, time and care was needed to achieve this consistently high duty factor.  Maintenance of the aging hardware as well as the time needed for the commissioning of new interferometer configurations both reduce the time available for observation. 
Simultaneously, ensuring that the data produced were of high quality and accurate calibration 
was made complicated by changes to the underlying instrumentation.  These
challenges inspired tools and analysis techniques that rapidly
evaluated the effect on sensitivity and likely origin of instrumental
noises to help inform commissioning improvements and maintain a sensitive,
stable detector.

The primary GW source candidates for GEO\,600
are gamma ray bursts (GRBs) or their progenitors, and supernovae (SNe), 
both of which produce short burst-like signals at high frequencies. 
The sensitivity of the detector to such sources depends not only on the
time-averaged level of the detector noise floor, but also on the degree of
Gaussianity of the detector noise. A putative GW burst signal must be
evaluated in the context of the background of instrumentally produced
transients that would trigger the same search. A focus was placed on the
development of monitors and figures of merit (FOMs) that track the
stationarity of the GEO\,600 data and make it easily accessible to the
instrument commissioners. One example is the transient-noise visualizer,
called Omegamap~\cite{Adams2015Costbenefit}, 
developed to provide information complementary to that encoded in the
amplitude spectral density. Whereas a spectrum assumes stationarity of
the data, the Omegamap time-frequency representation emphasizes the deviations
from that stationarity. The Omegamap is related to a normalized spectrogram,
but is one in which all FFT length choices are included and ranked by
significance, allowing the observation of transients of any
time-frequency character. 

Prompt and thorough accessibility to data quality monitoring and detector performance was realized through the development of an improved version of the overview web pages that had been used in the initial detector era.  These `summary pages'  are built upon a new universal architecture based on the computer cluster infrastructure used by
GW scientific collaborations for data analysis of the full detector
network data. This enhances the portability of monitor tools development and ensures functionality into the future. The summary pages organize and present monitors in a concise manner via
standard web servers. Month-long and daily summaries of the
detector's state are easily navigable and can be generated for any GW
detector~\cite{atlasDC2015}.
Other examples of new monitors that were developed during GEO-HF include the live noise budget presented in Sec.\,\ref{sec:challenges} as well as a real-time squeezing estimator. This vital connection between
the detector data quality and commissioning activities led to a novel
analysis framework for evaluating the cost-benefit of carrying out
particular commissioning activities rather than keeping the detector
in observation mode \cite{Adams2015Costbenefit}. 

\begin{figure}[tb]
\centering
\includegraphics[width=0.8\columnwidth]{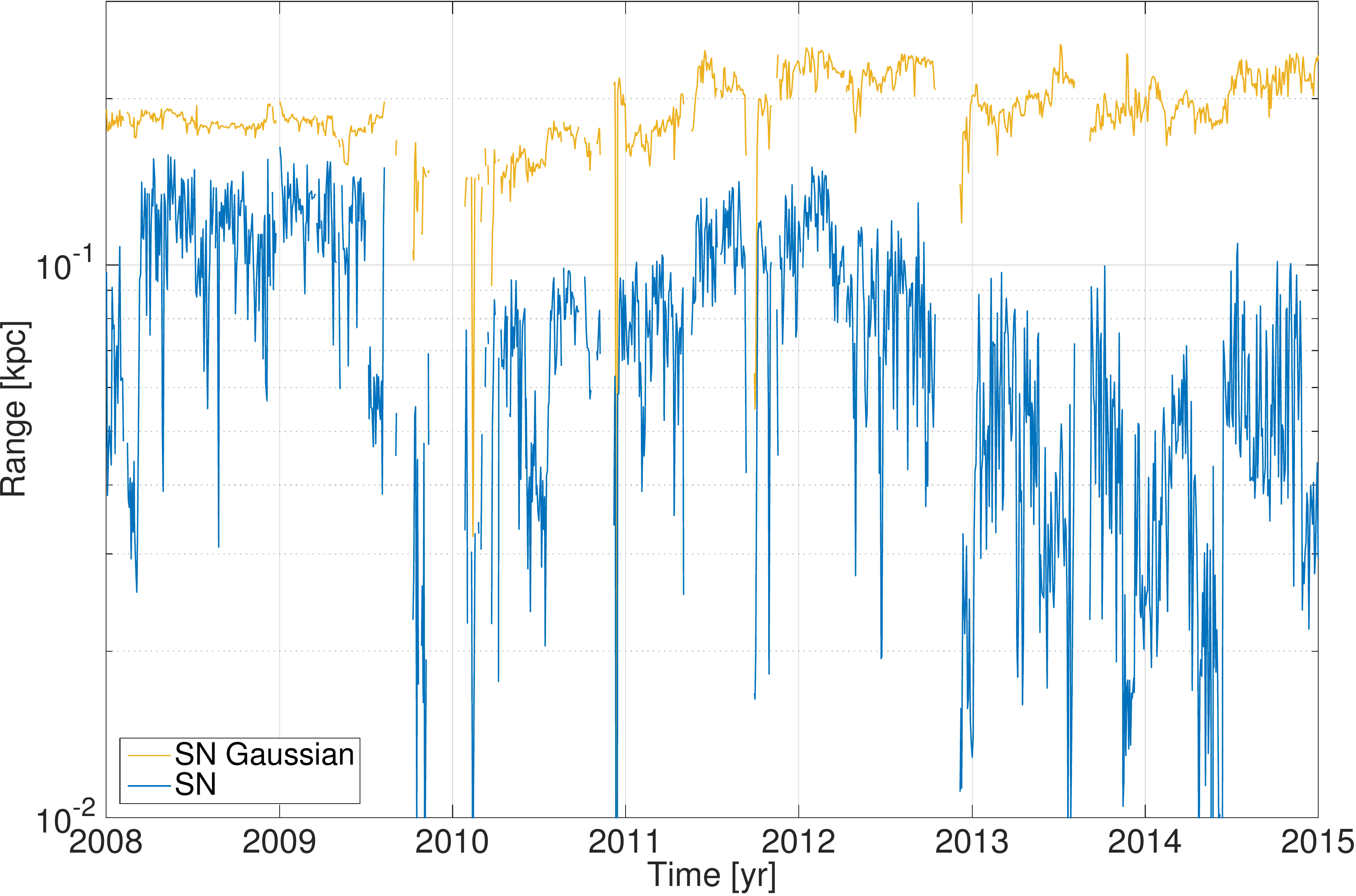}
\caption{Figures of merit tracking the astrophysical reach of GEO\,600 during GEO-HF to supernovae. The yellow trace (SN gaussian) shows the distance out to which GEO\,600 could detect a supernova based on the time-averaged detector noise floor. The blue trace (SN) takes into account not only the time-averaged noise floor, but also the glitchiness of the data.}
\label{fig:range}
\end{figure}


Figure~\ref{fig:range} shows the time series of two FOMs for the astrophysical
sensitivity of GEO\,600. They both present the range to which GEO\,600
could detect a supernova (SN). The dark yellow trace, similar to the
binary neutron star inspiral range commonly used by the global network
of detectors, is calculated from the time-averaged detector
noise floor weighted for this type of source. It assumes Gaussian
noise and uses a detection threshold with a SNR of 8. The blue trace
uses the same averaged noise floor but adds information from the
transient noise characteristics of the data. This non-Gaussian SN range is one of the most important new FOMs developed during GEO-HF. It 
allows for a more versatile estimate of possible analysis reaches and 
can be used to guide commissioning to focus on improving the
transient noise characteristics of a detector~\cite{Was2014Fixed}. 
This new FOM was adapted to sources such as GRBs 
or SNe, and it can be tailored 
for inspiral searches as well.
In order to assure
accuracy of these traces with respect to the astrophysics, calibration
of the data was provided through real-time parameter tracking and an
occasional absolute calibration~\cite{Affeldt2014Advanced, Leong2012New}.

During the course of the GEO-HF upgrades, the Gaussian SN range
improved by about 20\,\%.  At the end of the time period reported
here, this range reached 220\,pc, which reaches out to the star Betelgeuse.  
The large sensitivity increase seen above
2\,kHz in Figure~\ref{fig:h} is not reflected here because the SN ranges
integrate frequencies between 500\,Hz and 4\,kHz with a stronger emphasis
on the lower frequencies.  
The progression of the non-Gaussian SN range is much more variable and does not
reach above 150\,pc. It largely reflects the non-stationarity of the
data and highlights the difficulty of achieving data Gaussianity,
in particular with ongoing detector configuration changes and
commissioning.

\begin{figure}[tb]
\centering
\includegraphics[width=1.0\columnwidth]{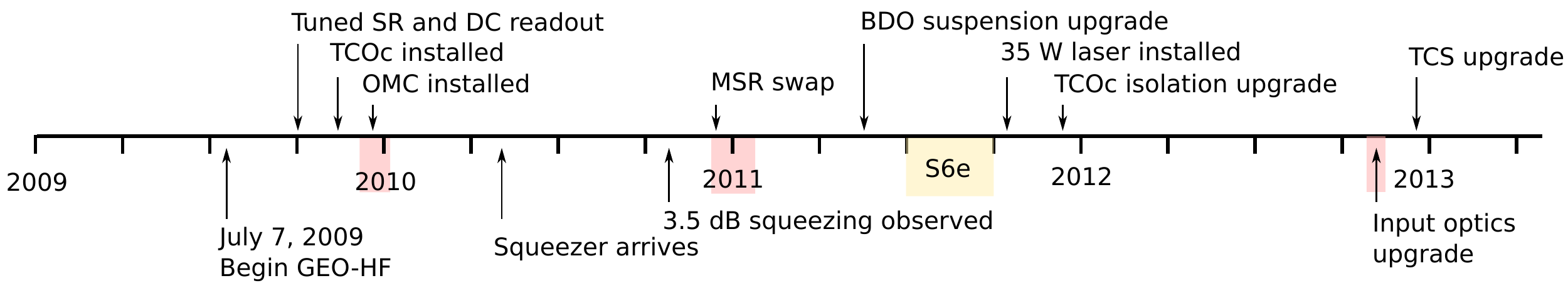}
\caption{Timeline of major events during the GEO-HF program. GEO-HF
  officially started when the Enhanced LIGO detectors began their S6
  science run on July 7, 2009. Tuned signal recycling (SR) together with DC
  readout were the first upgrade items to be implemented, followed by
  installation of the output mode cleaner (OMC), squeezing, signal 
  recycling mirror (MSR) swap,
  and finally, steps required to increase the laser power. During this time,
  upgrades of the output chain seismic isolation (BDO suspension) also
  took place to reduce technical noises that became more prominent. In all,
  during the GEO-HF period, 63\% of the time was spent collecting Astrowatch
  data, which includes a short science run with Virgo called S6e. TCOc:
  detection chamber containing the OMC; TCS: thermal compensation system.}
\label{fig:timeline}
\end{figure}

Several distinct commissioning epochs can be observed in
Figure~\ref{fig:range} and are summarized in a timeline in Figure \ref{fig:timeline}.
After implementing DC readout and the OMC in late 2009, both SN
ranges drop due to the fact that DC readout created an increased noise
floor and glitches, in part from the beam jitter coupling described in 
Section~\ref{sec:DCOMC}. 
To mitigate this, the output optics suspensions were improved in
May 2011. A small gap in the range data at this time is then followed by an
improvement in sensitivity as seen by both FOMs. 
Directly after
the output optics suspension upgrade and until March 2012, effort was put
into eliminating some of the transient noise features
at mid-frequencies from 100\,Hz to 1\,kHz.  The result of this work is
just barely visible in the non-Gaussian SN range.  
In 2012 several new glitch families began limiting the non-Gaussian SN range,
and a number of them have not been successfully tracked down. 
The severe drop in the
non-Gaussian SN range in November 2012 came after the input optics
upgrade. The new input mode cleaner has a lower finesse, and therefore provides less passive filtering of laser frequency noise, which resulted in
saturation of the laser frequency detection electronics. Fixing that is the improvement between
December 2012 and January 2013. In all, it is clear that it is essential to constantly expend commissioning effort to
maintain the astrophysical sensitivity by hunting for glitches and
other noise sources which are constantly arising.


To date, no GW detections have been reported with GEO\,600
or any other interferometric GW detector.
No interesting candidate events from
electromagnetic observations of nearby inspiraling neutron star binaries or
from galactic core-collapse supernovae have occurred during the Astrowatch period. 
Gamma ray bursts, however, triggered a coherent burst analysis for GWs associated with 129 GRBs between 2006 and 2011 using data from GEO\,600 and one of the LIGO or Virgo detectors. This was the first GW analysis performed using data from a GW detector using squeezed-light states to improve its sensitivity. No evidence for GW signals was found with any individual GRB in the sample or with the population as a whole \cite{LIGOScientificCollaborationandVirgoCollaboration2014Methods}.


\section{Summary and outlook}
\label{sec:summary}

The GEO-HF upgrade program started in 2009, and by today most of
the planned items have been successfully implemented.
The shot-noise-limited sensitivity has improved by up to a factor of 4,
and up to 3.7\,dB of squeezing have been achieved.
Work remains to be done reaching even higher squeezing levels and
potentially increasing the circulating laser power.
Besides these items GEO\,600 has also served successfully as a data-taking instrument
to fill the observational gap of larger observatories during their
upgrades to advanced detectors. Since 2008, more than 4 years worth of data
have been collected, allowing for serendipitous discovery in case of a nearby
GW source.
On top of this, GEO\,600 always served and serves as a testbed for novel
technology, making use of the fact that technologies can be tested
in an observatory environment.
A number of resulting contributions to the field have
been referred to throughout this paper, such as thermal compensation,
squeezing application and control, and software monitors to aid commissioning.

The LIGO detectors approach their first observational run late in 2015. Nevertheless there will be observational gaps over the coming years in the LIGO-Virgo
network. For the imminent future we therefore intend to operate GEO\,600 in a
similar fashion as to now, maintaining a reasonable fraction of observing time, 
and continuing to test technology. 
The large longer-term future of GEO\,600 still has to be determined.

\section*{Acknowledgements}
The authors are grateful for support from the Science and Technology
Facilities Council (STFC) Grant Ref: ST/L000946/1, the University of Glasgow in the UK, the
Bundesministerium f\"ur Bildung und Forschung (BMBF), and the state of
Lower Saxony in Germany. This work was partly supported by DFG grant
SFB/Transregio~7 Gravitational Wave Astronomy. This document has been
assigned LIGO document number LIGO-P1500140.

\section*{References}
\bibliographystyle{iopart-num}

\providecommand{\newblock}{}

\end{document}